\documentclass{iaus}
\usepackage{graphicx}

\newcommand{\mi}{$\mu$m}

\title[~~Dust in dwarf galaxies] 
{Dust in dwarf galaxies: The case of NGC 4214}

\author[U. Lisenfeld et al.]   
{U. Lisenfeld$^1$,
I. Hermelo$^1$,
 M. Rela\~no$^1$,
  R. Tuffs$^2$,
 J. Fischera$^3$,
B. Groves$^4$, 
\and 
C. Popescu$^5$ 
 }

\affiliation{$^1$Universidad Granada, Granada, Spain, email: {\tt ute@ugr.es, mrelano@ugr.es,israelhermelo@ugr.es} \\
[\affilskip]
$^2$Max-Planck-Institut f\"ur Kernphysik, Heidelberg, Germany, email: {\tt Richard.Tuffs@mpi-hd.mpg.de}   \\
[\affilskip]
$^3$Canadian Institute for Theoretical Astrophysics (CITA), University of Toronto, email: {\tt fischera@cita.utoronto.ca}  \\ [\affilskip]
$^4$Max-Planck-Institut f\"ur Astronomie, Heidelberg, Germany, email: {\tt brent@mpia.de}  \\ [\affilskip]
$^5$Jeremiah Horrocks Institute for Astrophysics and Supercomputing, University of 
Central Lancashire, Prestion, U.K., email: {\tt cpopescu@uclan.ac.uk}  }

\pubyear{2011} 
\volume{284}  
\pagerange{1--12}
\jname{The Spectral Energy Distribution of Galaxies}
\editors{R.J. Tuffs \&  C.C.Popescu, eds.}
\begin{document}

\maketitle

\begin{abstract}
We have
carried out a detailed modelling of the dust heating and
emission in the nearby, starbursting dwarf galaxy NGC 4214.
Due to its proximity and the great wealth of data 
from the UV to the millimeter  range (from GALEX, HST, {\it Spitzer},
Herschel, Planck and IRAM) it is possible to separately model the 
emission from HII regions and their associated photodissociation regions (PDRs) and 
the emission from diffuse dust. 
Furthermore,  most model parameters can
be directly determined from the data leaving very few
free parameters.
We  can fit both the emission from HII+PDR regions
and the diffuse emission in NGC 4214 with these models
with "normal" dust properties and realistic parameters.
\end{abstract}

\firstsection 
\section{Introduction}

The dust spectral energy distribution (SED) of dwarf galaxies frequently  show differences 
to those of spiral galaxies. The two main differences are:
(i) a relatively low emission at 8\mi , most likely due to 
a lower PAH content at low metallicities (e.g. Draine et al. 2007, Engelbracht et al. 2008),
and (ii)  a submillimeter (submm) "excess" which has
been found in the SED of many starbursting, low-metallicity galaxies (
Lisenfeld et al. 2002, Galliano et al. 2003, 2005, 
Bendo et al. 2006, Galametz et al.
2009, 2011, Israel et al. 2010, Bot et al. 2010).
Different reasons  have been suggested to explain this excess:

(1) A large amount of cold ($<$ 10 K) dust (Galliano et al. 2003, 2005, Galametz 2009, 2011). 
However, extraordinarly large dust
masses are needed for this explanation and it is unclear, how these large amounts
of cold dust can be shielded efficiently from the interstellar radiation field (ISRF).

(2)  A low value of the dust emissivity spectral index of $\beta = 1$ in the submm.
Different dust grains have been suggested that could be responsible for this,
from very small grains (Lisenfeld et al. 2002), fractal grains (Reach et al. 1995)
to amorphous grains (Meny et al. 2007).

(3) Spinning grains (Ferrara \& Dettmar 1994; 
Draine \& Lazarian 1998). Bot et al. (2010) showed that this grain type could
explain the submm and mm excess in the Large and Small Magellanic Cloud.

In order to  interpret the dust SED of a galaxy and understand the differences of
the SED of dwarf galaxies, a physical model is needed,  based on realistic
dust properties and taking into account  the heating and emission 
 of dust immersed in the wide range of  ISRFs. 
 Ideally, radiation transport in a realistic geometry
should be done, but is often difficult due to the complex geometry and
large number of parameters. Models can generally be  classified into three broad groups:
(1) modified blackbody fits, which are too simple to describe reality correctly but give a first idea
of the range of dust  temperatures,
(2) semiempirical models that try, in a simplified way, to
describe dust immersed in a range of  different radiation fields
 (e.g. Dale et al. 2001, Draine et al. 2007, Galametz et al. 2009, 2011, da Cunha et al. 2008) and
(3) models that include full radiation transfer (e.g. Popescu et al. 2011 for  spiral galaxies, Siebenmorgen \& Kr\"ugel 2007 for
starburst galaxies) which are the most precise description of a galaxies if all parameters, including the geometry, is known.

In the present work we model the dust emission of the nearby ($D=2.9$ Mpc) 
starbursting dwarf galaxy NGC 4214.
Due to its proximity and the large amount of ancillary data  it is possible to apply physical
models that take into account the full radiation transfer  and constrain
their input parameters.  
In the present contribution we describe the general outline of our work and the results, whereas a more
detailed description of the available data, the data reduction and the determination of the model input
parameters
is presented in the contribution of Hermelo et al. in this volume.

\section{The observed SED of NGC 4214}

NGC 4214 is an irregular galaxy, dominated by two bright, young star-forming (SF) regions in its center
(called NW and SE in this work).
A great wealth of data is available for this object, ranging
from GALEX ultraviolet (UV), Hubble Space Telescope (HST) UV to infrared (IR) images,
Spitzer IRAC and MIPS, Herschel SPIRE,  IRAM MAMBO at 1.2mm, as well as Planck detections at 350,
550 and 850 \mi . Furthermore the galaxy has been mapped in HI as part of the THINGS
project (Walter et al. 2008) and has been observed with OVRO in CO(1-0) (Walter et al. 2001).

The large amount of  data, most of them at a high spatial resolution, allows to
determine the dust SED separately for the emission from the SF regions SE and NW,
where the dust is heated by nearby massive stars, and the diffuse dust heated by
the general interstellar radiation field.  Figure 1 shows the observed dust SED of the individual SF
regions   (top) and of the diffuse medium (bottom),  determined
by subtracting the emission of both SF regions from the total dust emission.



\section{The models}

We  separately modelled the emission from the massive SF regions and the diffuse dust emission.
We use the model of Groves et al. (2008) that describes the emission from an HII region together with its
surrounding PDR for the SF regions. It describes the  luminosity evolution of a star
cluster of mass $M_{\rm cl}$ from stellar population synthesis, and incorporates  the expansion of 
the HII region and PDR due to the mechanical energy input of stars and SNe.
The dust emission from the HII region and surrounding PDRs is calculated from radiation transfer.
The main parameters in this model are: (1) metallicity, (2) age of the cluster, (4) external pressure, (3) 
compactness parameter, $C$, which
parametrizes the heating capacity of the stellar cluster and depends on $M_{\rm cl}$ and the external pressure
of the ambient medium,
(5) column density of the PDR, $N_{\rm H}$, and  (6) covering factor, $f_{\rm cov}$, defining which fraction of the HII region is surrounded by the PDR.
Due to the great wealth of data and previous studies carried out, we can constrain most parameters (all except
$N_{\rm H}$) very tightly
from observations (see Hermelo et al.).

In order to fit  the diffuse emission, we applied the templates from Popescu et al. (2011), calculated including  the full radiation transfer for a disk galaxy. The model consists of
two exponential stellar disks, describing old and young stars, respectively,  together
 with their associated dust disks. Two additional components of the model can be neglected in our
case: a  bulge which is absent in NGC 4214 and 
individual young SF regions which we subtracted from the SED and
modelled separately with the model of Groves et al. (2010).
The main parameters of the diffuse model are: (1) the central face-on opacity $\tau_B$, (2) the 
SF rate (SFR) of the young
stellar population, (3) the SFR of the old stellar population and  (4) the exponential scale-length, $h_s$, of the stellar emission. All these parameters can be tightly constrained by the observations (see Hermelo et al.).

\section{Results and conclusions}

\begin{figure}[b]
\begin{center}
 \includegraphics[width=6.cm]{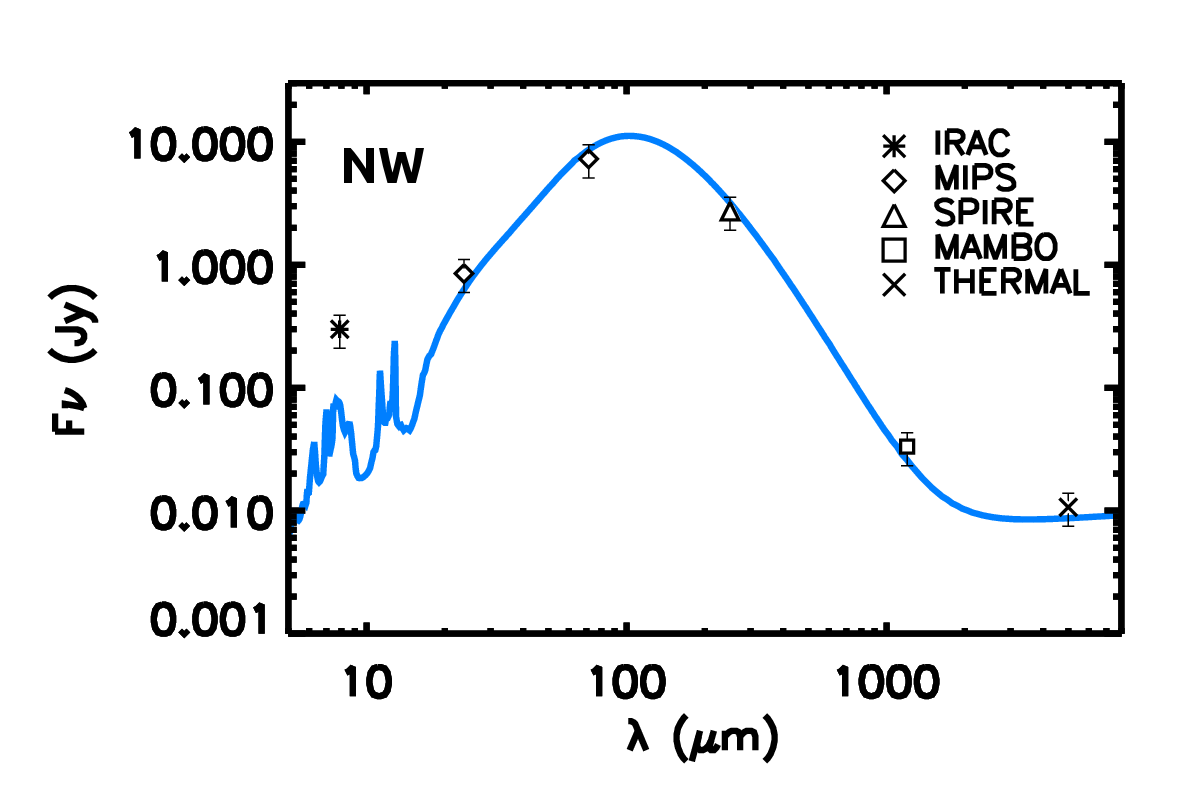} 
 \includegraphics[width=6.3cm]{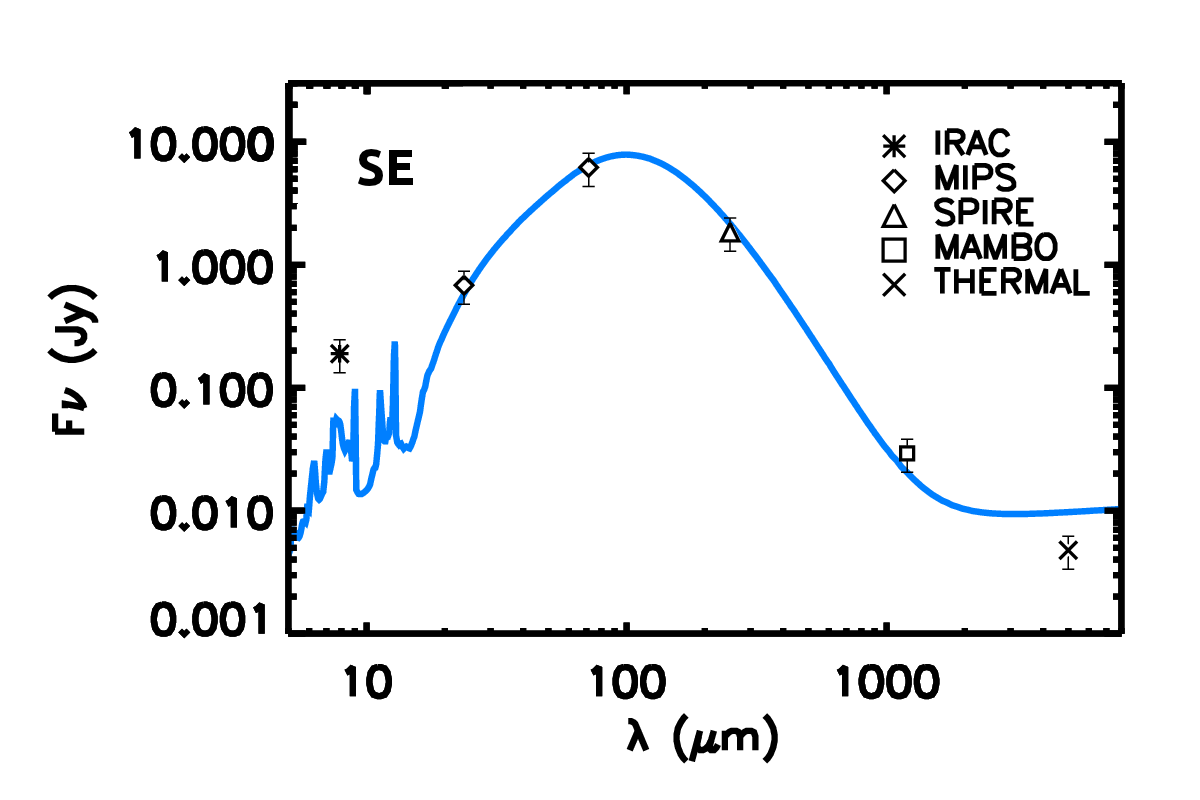} 
 \includegraphics[width=6.cm]{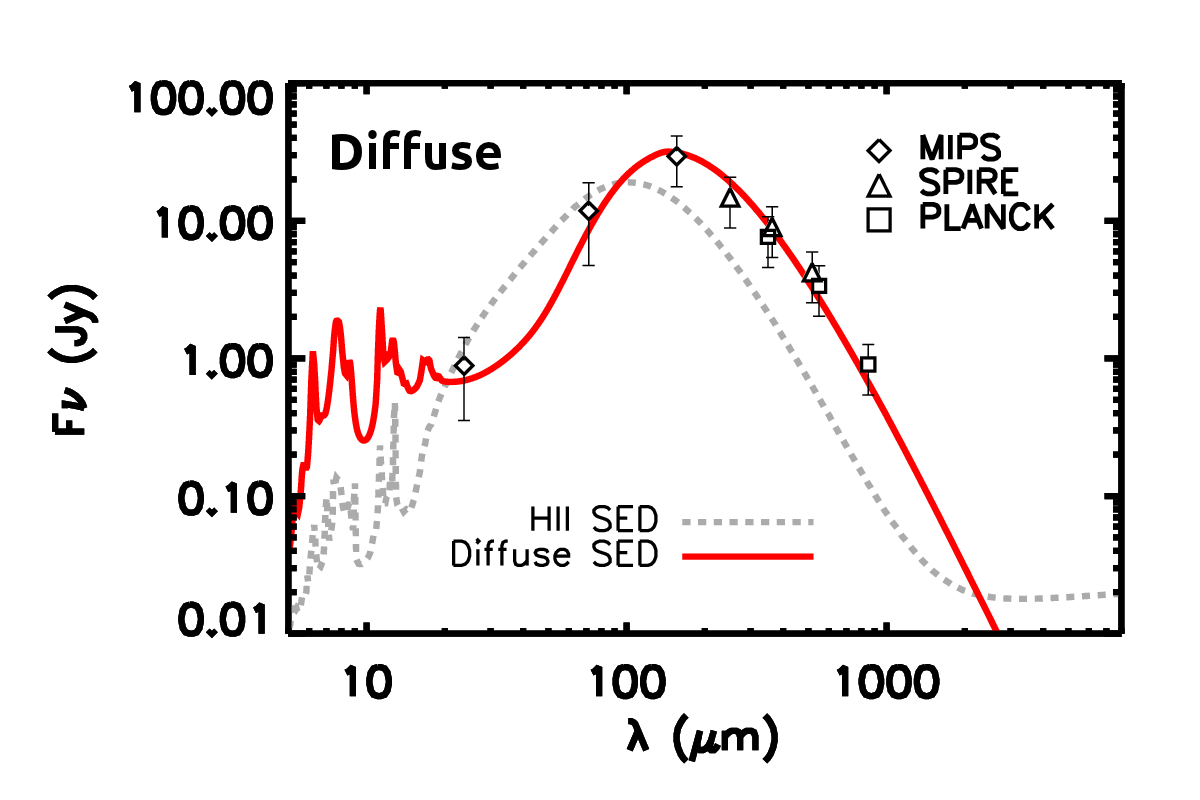} 
 \caption{\textit{Top:} The dust emission of the HII regions NW (left), SE (right) and the best-fit model
 for the parameter range determined by the observations. \textit{Bottom:} The same for the 
 diffuse emission. For illustration, the sum of the fit to NW+SE is also included. 
 }
   \label{fig1}
\end{center}
\end{figure}

In Fig. 1 we show the fits of the models to the data, separately for both SF regions and 
for the diffuse emission. The parameters were chosen within the tight constraints given by the
observations.
Note that for the diffuse emission the flux level of the dust emission is not  free but is
 fixed by  the observationally measured SFR.
 
 The model by Groves et al. (2008) for the region NW fits all data points longwards
 of 10 \mi\ within the errors. The fit for the SE region is good,
 but  underestimates slightly the long wavelength data at 1.2mm and  overestimates
 the thermal radio emission.  Here, a better fit could be achieved for  
 ages higher than those estimated for the central clusters (4.5 instead of 3.5 Myr).
 This could indicate that the dust  within our apertures 
 is not only heated by the central clusters but
 also by an older stellar population.
 The model underestimates the emission at 8\mi , the reasons for this still need
 to be investigated.
 
We achieve an excellent fit for the diffuse dust emission. However, we had to assume a 
larger horizontal scale-length (980-1265 kpc instead of the observed 800 kpc) in order to produce the
correct dust temperature.
A possible reason for this discrepancy is  that the vertical scale height of the
stellar light in this irregular galaxy  is higher than the model value for normal spiral galaxies adopted  
in  Popescu et al.  (2011).

We can calculate the dust mass from eq. (44) of Popescu et al. (2011)
($M_{dust} =   \tau_B \times h_s^2 \times 0.99212$ pc$^{-2} M_\odot$).
With our values of $h_{\rm s} = 980$ kpc and $\tau_B = 1$ we derive a dust mass
of $9.7 \times 10^5$ $M_\odot$. Together with $M_{\rm HI} = 4.1 \times 10^8$ $M_\odot$
(Walter et al. 2008) and $M_{\rm H2} = 5.1 \times 10^6$ $M_\odot$ (Walter et al. 2001, obtained
with a Galactic conversion factor) we derived a gas-to-dust mass ratio of 470. This
is about three times the Galactic value which is realistic for a galaxy with a $\sim$  three
times lower metallicity.

In conclusion, we find that we obtain  a generally good agreement between model
predictions, based on standard dust properties, and the data. There are no indications
for a submm excess. NGC 4214
is a special case because the wealth of ancillary data and its proximity allowed
to (i) determine the SED of the dust emission separately for the dust in SF regions
(HII region + PDR) and the diffuse dust and (ii) constrain most model input parameters
independently from observations.
The fact that we achieved a good agreement between models and data shows 
that this approach is a fruitful way to try to understand the SED of nearby dwarf
galaxies and to find the origin of peculiar features as e.g. the submm excess.

\end{document}